# Proliferation of neutral modes in fractional quantum Hall states

Hiroyuki Inoue, Anna Grivnin, Yuval Ronen, Moty Heiblum, Vladimir Umansky and Diana Mahalu

Braun Center for Submicron Research, Department of Condensed Matter Physics, Weizmann Institute of Science, Rehovot 76100, Israel

**The fractional quantum Hall effect (FQHE) is a canonical example of a topological phase in a correlated 2D electron gas under strong magnetic field[1,2]. While electric currents propagate as chiral *downstream* edge modes[2-4], chargeless *upstream* chiral neutral edge modes were recently observed only in *hole-conjugate* states (states filling $v$, $n+½<v<n+1$, with $n=0,1,2,…$), and in the even denominator state $v=5/2$[5-9]. It is believed that spontaneous 'density reconstruction' near the edges of the 2D gas, leads to multiple counter propagating edge channels, being separated from each other by incompressible strips. Unavoidable disorder induces inter-channel tunneling; accompanied by Coulomb interaction it renormalizes the multiple edge channels to a downstream charge mode and upstream neutral edge mode(s), while maintaining the parity requirements (e.g., the heat conductance along the edge) dictated by the bulk[10-17]. Here, we report of highly sensitive shot noise measurements that revealed unexpected presence of neutral modes in a variety of non-hole-conjugate fractional states. As already reported previously[5], we did not observe neutral modes in any of the integer states. In addition to the upstream neutral edge modes, we were surprised to find also neutral energy modes that propagate through the incompressible bulk. While along the edge, density reconstruction may account for the edge modes, we are not aware of a model that can account for the bulk modes. The proliferation of neutral modes, in every tested fractional state, changes drastically the accepted picture of FQHE states: an insulating bulk and 1D chiral edge channels. The apparent ubiquitous presence of these energy modes may account for decoherence of fractional quasiparticles - preventing observation of coherent interference in the fractional regime.**

Charge propagation in the quantum Hall effect regime takes place near the edge of the 2D electron gas in form of *downstream* chiral edge modes (with chirality determined by the direction of the magnetic field); while the bulk is incompressible (gapped) and thus does not contribute to the transport[1-4,18-19]. In the integer regime, with *v* the integer filling factor (being the ratio between the number of electrons to the number of flux quanta, $\phi_0=h/e$), charged excitations are electrons while in the fractional regime (say, *v*<1), they are fractionally charged quasiparticles (or quasiholes)[20-23]. However, as was predicted in the 90's[10-11], and recently demonstrated[5-9], this picture is, in general, too simple. For, the so called, hole-conjugate states of the FQHE regime, edge *reconstruction* of the electron density is expected to take place at the physical edge of the 2D gas; with counter-propagating electron and hole edge currents. Since such edge currents were never observed[12], tunneling between these edge channels (due to disorder), accompanied by Coulomb interaction, had been proposed to renormalize the 'multiple current channels', leading to a *downstream* chiral current mode and *upstream* (counter propagating to the charge modes) neutral (chargeless) mode(s)[13-15] – as was indeed recently detected in hole-conjugate FQHE states[5-9].

Following a significant effort to develop a sensitive heat sensor in a thinly polished substrate[24], our group detected the presence of neutral modes in *v*=2/3, 3/5, 5/3, 8/3 and 5/2 via measuring noise, with zero net current, resulting from impinging the modes *upstream* on a partly pinched quantum point contact (QPC) constriction, or, alternatively, by on a non-ideal ohmic contact[5-7]. The measured noise may result from the fragmentation of the neutral quasiparticles (it is instructive to look at them as electrical dipoles) by a partitioning barrier[5], or, alternatively, due to local Joule heating of the barrier[7]. Support to the latter mechanism was also provided by observing thermal broadening of conductance peaks[8,9] or the generation of net thermoelectric current[9] - both in quantum dots (QDs). Though one can find predictions for edge reconstruction also in integer fillings[25] and also in particle-like fractional states[26-27], they were not observed before. Here, we report on findings of chiral upstream neutral modes, again via sensitive noise measurements, in particle-like fractional states. Moreover, and even more surprising, energy flow through the bulk was also found in all the tested states. Note, that bulk heat transports in $v=1^8$ and in $v=4/3^{28}$ were recently reported.

Excited neutral modes, carrying energy, can be excited by two possible sources (see schematic illustration in Fig. 1 (also, S1 in Supplementary Section): (*i*) A partly pinched QPC, which partitions downstream current $I_C$ arriving from $S_C^7$, and (*ii*) The *hot-spot* at the upstream side of a biased ohmic contact $S_N^5$, fed by current $I_N$. In the following experiments, neutral modes were converted to downstream current fluctuations, $i^2$, with zero net current, in standard ohmic contacts[7]. Voltage fluctuations, $v^2=i^2/v^2G_0^2$ (independent of the QPC transmission)[29], were amplified by a cryogenic preamplifier; placed along the downstream edge at $A_{CE}$, along the upstream edge at $A_{NE}$, or connected to the bulk at $A_{NB}$. The presence of neutral modes was verified by measuring shot noise in two configurations (Fig. 1): (*i*) Partitioning $I_C$ by a QPC and measuring the resultant upstream noise at $A_{NE}$ and the bulk noise at $A_{NB}$ - named C→N; and (*ii*) As in (*i*), but the noise at $A_{CE}$ was measured in the presence of a neutral mode arriving upstream from $S_N$ - named C+N. The latter measurements allowed measuring the charge of the partitioned quasiparticles and their temperature[29,30] in presence of an impinging neutral mode. It is worth reminding the reader of the 'standard' expression of excess noise (shot noise above the thermal contribution), due to stochastic partitioning (transmission *t*) of a noiseless beam of charge $e^*$ quasiparticles, $S_i = 2e^*It(1-t)(\coth x - x^{-1})$, with $x = e^*Iv^{-1}G_0^{-1}/2k_BT$, where $S_i$ is the 'zero frequency' noise spectral density, $k_B$ the Boltzmann constant, and $I$ the impinging current[29,30].

We start with filling $v=2/3$ of the first spin-split Landau level, where the presence of upstream neutral modes had been already established[5,7-9]. We fabricated a device with short propagation distance of the neutral modes, 3.5μm & 7μm, in order to enhance the measured shot noise. At the center of the $v=2/3$ conductance plateau, $B=6.7$T, with the state fully polarized, the neutral mode was measured in the C→N configuration, with -4nA≤$I_C$≤4nA (see Fig. 1). Another configuration, named N→N, relied on exciting the neutral mode by the 'hot spot' at the upstream side of $S_N$ by driving current $I_N$. The neutral mode was partly transmitted by the upstream QPC and generated currentless noise at $A_{NE}$ (S2 at the Supplementary Section). In all the measurements, only a negligible amount of net current (<0.5pA, via through the bulk) was detected at $A_{NE}$ & $A_{NB}$; assuring a negligible contribution to the noise (if carrying poissonian noise (see also Fig. 6b), one would expect $S_i=2eI_{bulk}\sim3\times10^{-31}A^2/Hz$). A monotonically increasing (currentless) excess noise with $I_C$ was measured upstream along the edge (at $A_{NE}$) and directly through the bulk (at $A_{NB}$); both exhibiting $\sim t(1-t)$ dependence (see Figs. 2a & 2b). Two main observations are apparent from the data: (*i*) The excess noise along the edge in $A_{NE}$, exhibiting a slight negative curvature, is some ten times higher than that in $A_{NB}$; (*ii*) The excess noise through the bulk in $A_{NB}$ exhibits a slight positive curvature. Note that the integrity of the state is maintained by keeping the excitation voltage in $S_C$ (150μV) substantially lower than the presumed energy gap of $v=2/3$ (~780μV)[31].

In the C+N configuration at filling $v=2/3$, we measured the noise at $A_{CE}$ in the range -4nA≤$I_C$≤4nA and 0nA≤$I_n$≤4nA (Fig. 3). Two contributions to the downstream excess noise at $A_{CE}$ were measured (Fig. 3a): one due to partitioning of $I_C$ in the presence of $I_N$, and the other due to fragmentation of the neutral mode arriving from $S_N$ (or excess heating). Assuming these are uncorrelated contributions, we can easily extract the temperature and charge of the partitioned quasiparticles (note, that detailed mechanism of neutral-charge interaction in the QPC is not known)[5-7]. At $I_N=0$, with increasing $I_C$ the excess noise behaves binomial (above expression), with the partitioned quasiparticle charge $e^*\sim2e/3$ at temperature $T=25$mK (Fig. 3b). As $I_N$ increased from 0 to 4nA, the contribution of the neutral mode to the noise increased, and the partitioned quasiparticle charge gradually decreased from $e^*\sim2e/3$ to $e^*\sim0.4e$. Similarly, the temperature of the partitioned quasiparticles increased to ~140mK at $I_N=4$nA, as shown in Fig. 3c; agreeing qualitatively with previous measurements[5,32]. We note that such temperature increase barely influenced the transmission $t$ of the QPC.

We move now to $v=1/3$, where neutral modes were not observed before; and while not-forbidden theoretically, their presence was not expected. Employing first the C→N configuration, with the magnetic field tuned to the center of the conductance plateau, $B=13.5$T. As shown in Figs. 2c & 2d, excess upstream noise was observed, accompanied also by noise in $A_{NB}$. Though considerably weaker than in $v=2/3$, the excess noise was easily detected (due to the shorter propagation length and the extremely 'quiet' preamplifiers), exhibiting also $\sim t(1-t)$ dependence; with the noise along the edge some eight time stronger than that through the bulk. As in the $v=2/3$ case, the excitation voltage is maintained much smaller than the gap energy[31].

Measurements at the C+N configuration followed at $v=1/3$ (Figs. 4). The excess noise followed the familiar dependence on current; however, the partitioned quasiparticle charge varied with bias (even though $t$ was constant). Concentrating on the small bias regime ($I_C<2$nA), the partitioned quasiparticle charge (at $I_N=0$) was $e^*\sim0.78e$ (Fig. 4b). This is not surprising, as quasiparticles bunching is often found when $t$ is not nearly unity[33]. As $I_N$ increased, the quasiparticle charge decreased to $e^*=0.56e$, accompanied by temperature increase of the partitioned quasiparticles to $\sim85$mK (Fig. 4c).

Having established the presence of currentless neutral modes in $v=2/3$ and also in the unexpected $v=1/3$ states, we also tested other particle-like states, such as $v=2/5$ and $v=4/3$. In both cases, only the current carried by the inner channels (2/5-1/3, and 4/3-1) was partitioned by the QPC in the C→N configuration. The results are qualitatively similar to those in $v=1/3$ states. (see S3 in Supplementary Section for C→N configuration at $v=2/5$). Though it cannot be fully discounted[25], integer fillings did not have any measurable noise in the C→N configuration; even with $I_C=4$nA and background noise level $<5\times10^{-31}$A$^2$/Hz (see S4 in Supplementary Section).

Since spurious thermal effects may contribute to the excess noise, measurements in the C→N configuration were repeated at temperature T=50, 100, and 180mK at $v=2/3$ and $v=1/3$. Strong reduction of the excess noise at $A_{NE}$ and $A_{NB}$ was observed (Fig. 5). Such noise suppression indicates: (*i*) The minute bulk current, which increased slightly with temperature (since $R_{xx}$ increased), does not contribute to the excess noise; (*ii*) The decay of the excess noise with temperature conforms qualitatively with the expected shortening of the decay length with temperature[13,14]; and (*iii*) The bulk mode seems to decay much faster in comparison with the edge mode. Having both similar propagation lengths, it may suggest that the modes are of two different natures.

In order to verify further the presence of upstream neutral modes in particle-like states, we employed a different geometry, which allowed continuously tuning of the filling fraction in the bulk via top gates; albeit not separating the bulk and edge contributions (see Fig. 6a for the schematics). We employed the C→N configuration ($I_C$=2.5nA); however, instead of partitioning the edge channel with a QPC as above, it was partitioned at the interface between two filling factors: arriving from $v_L$=1 region and partitioned into a variable $v_R$<1 region – with an assumed 'hot spot' at the partitioned region (see Fig. 6a). The length of the interface between TGL and TGR was 3μm and the total path length to the amplifier contact was 8μm. The upstream neutral mode was expected to propagate in the $v_R$<1 region, with noise produced in ohmic contact $A_{NE}$. In Fig. 6b we plotted the measured noise at $B$=7T as function of the gate voltage on the right top gate $V_{TGR}$ as well as the undesirable bulk current $I_{bulk}$ and it presumed Poissonian noise $2eI_{bulk}$. Evidence of upstream noise is clear in all fractional fillings between $v$=2/3 and $v$=1/3. Note that while the bulk current goes up in the compressible regions, the noise goes down – as was described above. As before, the noise in particle-like states is considerably smaller than in $v$=2/3 with vanishing bulk current in both cases. Varying the density of the bulk and adjusting correspondingly the magnetic field, we also performed noise measurements at constant filling fractions, with the results shown in some details in the Supplementary Section (S5).

The discovery of such proliferation of neutral energy modes, changes the accepted view of current and energy transport by chiral edge channels and the bulk in the FQHE. Though only a few fractional states had been tested here, it is highly likely that most fractional states, being particle-like or hole-conjugate-like, harbor a much more complex edge structures. Moreover, the bulk, being incompressible, seems to be active energetically. Since the topological properties of the bulk dictate a conserved quantum thermal conductance for each state (*e.g.*, zero for $v$=2/3; one for $v$=1/3)[15], it imposes fundamental differences between particle-like and hole-like states. Even for a sharp confinement potential (compared to magnetic length), the $v$=2/3 state must consist of two oppositely propagating channels, while the $v=1/3$ must not to reconstruct. Moreover, rather than having only two oppositely propagating edge channels in the $v=2/3$ state, for a shallow enough confining potential, an additional puddle of $v=1/3$ state can nucleate on the edge (thus adding a pair of counter-propagating 1/3 channels)[34,35]. In fact, our measurements were performed with 'gate defined' 2D structures (see Fig. 1), and thus, the electron density decreased gradually towards the edge – encouraging such edge reconstruction. Hence, such edge reconstruction might be eliminated in the latter, by sharpening the profile of the density near the edge, and thus affecting the presence and the nature of the neutral modes.

How can one envision the likely edge reconstruction in $v=1/3$ state? Very much like the parameters that play role in the $v=2/3$ state, a competition among the confinement potential, magnetic length, and the Coulomb interaction, dictate the electron distribution near the edge[34,35]. The electron density should, at first, decrease monotonously towards the edge, leading to a downstream channel. Subsequently, two (at least) counter-propagating channels must emerge: one upstream due to an increased electron density, and one downstream due to the density dropping to zero at the depletion region near the edge. Yet, the net conductance must be $G_0/3$; necessitating an upstream neutral mode. It is unlikely that the two added channels should be of $v=1/3$ character; since the original 1/3 channel and the added upstream one will easily localize due to disorder[36]. Hence, the added two channels may be of lower filling, such as $v=1/5$, or, alternatively, a higher one, such as $v=1$. This type of reconstruction may depend on details and should be confirmed experimentally at each fractional filling.

The presence of bulk modes may indicate bulk-edge coupling, even though some of the features and temperature dependence are different[28]. We are not aware of any prediction of such bulk energy transport, specifically in the fractional regime, which may account for our observation. The immeasurably small noise in $A_{NB}$ in the integer regime likely eliminates energy transport via crystal phonons emerging from the region of potential imbalance (ohmic contact or QPC)[28].

The proliferation of energy modes, especially those excited by partitioning of current in a QPC, may explain the difficulty in observing interference of fractional charges in interferometers that are based on charge partitioning. Drawing energy from the charged quasiparticles in the partitioning process, the modes may serve as sources of decoherence – preventing studying the statistics of fractionally charged quasiparticles.

## METHODS SUMMARY

For the noise measurement, a voltage source with 1GΩ resistor at its output fed the DC currents and the generated noise signals at amplifier contacts, filtered with an LC circuit tuned to ~800 kHz, were amplified first by a cooled, home-made, preamplifier with voltage gain 11 and subsequently by a room temperature amplifier (NF-220F5) with voltage gain 200. The amplified signal was measured by a spectrum analyzer with the bandwidth of 10 kHz. In order to monitor the net current reaching the amplifiers $A_{NE}$ and $A_{NB}$, 5µV$_{RMS}$ at the resonant frequency on top of the DC currents was applied from the sources and measured at amplifiers with the bandwidth of 30Hz. For the differential conductance measurement to characterize the QPC, we sent 0.5µV$_{RMS}$ at the resonant frequency along with DC currents and measured at $A_{CE}$ with the bandwidth of 30Hz.

## Acknowledgements

We thanks Y. Gefen and Y. Meir for useful discussions. We acknowledge the partial support of the Israeli Science Foundation (ISF), the Minerva foundation, the U.S.-Israel Bi-National Science Foundation (BSF), the European Research Council under the European Community's Seventh Framework Program (FP7/2007-2013)/ERC Grant agreement No. 227716, and the German Israeli Project Cooperation (DIP).

# Figures

**Figure 1. | Schematics of the experiment.** The device was fabricated on a GaAs/AlGaAs heterostructure, with 2DEG embedded 130nm below the surface whose carrier density is $1.0\times10^{11}$cm$^{-2}$ and the dark mobility is $5.1\times10^{6}$cm$^{2}$/Vs at 4.2K. The light blue colored region is the mesa embedding the 2DEG. The yellow pads are standard NiAuGe ohmic contacts, grounds (G), amplifiers ($A_{CE}$, $A_{NE}$ &$A_{NB}$), and sources, $S_C$ and $S_N$, driving charge current $I_C$ (blue thick arrows) and neutral current $I_N$ (red thick arrow). The grounded contacts were tied directly to the cold finger of the dilution refrigerator. Heating the mixing chamber raised the temperature of the electrons. The QPC (green lines) was formed by negatively biased split-gates (10nm Ti/30nm Au) with a 660nm wide opening, controlling the transmission probability $t$. The distance between $S_N$ and the QPC was 3.5μm. Each amplifier contact was followed by a resonant circuit (peaking ~800kHz) and a cryogenic voltage amplifier. $A_{NE}$ sits at 7μm along the edge upstream from the QPC, measuring the upstream neutral edge mode solely. $A_{NB}$ is located at 8.5μm over the mesa from the QPC, measuring the bulk mode solely. $A_{CE}$ was to measure the noise in the downstream charge mode. Thin arrows are unbiased channels. We employed two configurations: (*i*) C→N, driving only $I_C$, partitioned at the QPC and exciting the neutral modes if they were to exist, and detecting their presence by measuring the noise at $A_{NE}$ and $A_{NB}$ simultaneously and (*ii*) C+N, driving both $I_C$ and $I_N$ and measuring the noise at $A_{CE}$. The presence of the neutral modes can modify the effective charge and temperature of tunneling quasiparticles.

**Figure 2. | C→N noise at $v$=2/3 and 1/3 for several QPC transmissions.** The configuration C→N allows us to detect the neutral modes in the upstream edge and over the bulk separately. **a**, In $v$=2/3, the C→N noise at $A_{NE}$ for -4nA$\leq I_C \leq$4nA with transmissions $t$=0.65, 0.90 and 1.00. As $|I_C|$ increases, the noise also increases monotonically. The magnitude of the noise approximately go as $t(1-t)$. **b**, In $v$=2/3, the C→N noise at $A_{NB}$ measured simultaneously with $A_{NE}$. The magnitude of the noise in $A_{NB}$ is about ten times smaller than that in $A_{NE}$. **c**, In $v$=1/3, the C→N noise at $A_{NE}$ for -4nA$\leq I_C \leq$4nA with transmissions $t$ = 0.64, 0.72 and 1.00. The noise again increases

monotonically with $|I_C|$. The magnitude of the noise follows nearly $t(1-t)$. **d**, In $\nu=1/3$, the C→N noise at $A_{NB}$ measured simultaneously with $A_{NE}$. The magnitude of the noise in $A_{NB}$ is about 7 times weaker than that in $A_{NE}$. No kink that may originate from the bulk gap was seen in any cases above.

**Figure 3. | C+N noise at $\nu=2/3$.** The influence of the $I_N$ on the partitioning the charge current $I_C$ also provides indications of the presence of the neutral modes. **a**, The C+N noise at $A_{CE}$ for -4nA$\leq I_C \leq$4nA with $t=0.72$ was measured in the presence of 0nA$\leq I_N \leq$4nA in steps of 0.5nA. While the traces are merging at high $I_C$ region, one can clearly see an increase in the cut at $I_C=$0nA, which is consistent with the C→N measurement and shows the N→C process is also present. Note that the transmission as a function of $I_C$ was barely affected by the presence of $I_N$. **b**, The fitted temperature as a function of $I_N$ shows the rise of the electron temperature, from about 20mK up to 140mK, with increasing $I_N$. **c**, The extracted effective charge as a function of $I_N$ shows the suppression of the effective charge, approximately from $2e/3$ down to $e/3$, as reported before.

**Figure 4. | C+N noise at $\nu=1/3$.** Another verification of the neutral mode in $\nu=1/3$ in the C+N configuration. **a**, The C+N noise at $A_{CE}$ for -4nA$\leq I_C \leq$4nA with $t=0.83$ was tested with 0nA$\leq I_N \leq$4nA in steps of 1nA. A distinct increase in the low $|I_C|$ region ($\leq$2nA) showed up, again being consistent with the C→N measurement and witnessing the N→C process. Note that $I_N$ had negligible effect on the transmission as a function of $I_C$. **b**, The electron temperature as a function of $I_N$ rose from about 25mK up to 80mK with increasing $I_N$. **c**, The effective charge as a function of $I_N$ decayed approximately from $0.8e$ down to $0.5e$, relaxing the quasiparticles bunching, similar to a previous study.

**Figure 5. | C→N noise at $\nu=2/3$ and 1/3 at elevated temperatures.** We repeated the measurement of the C→N noise at 50, 100 and 180mK for $\nu=2/3$ and 1/3 at both amplifiers. In all the cases, the noise decayed with raising the temperature. The bulk noise weakened

faster than the edge noise, indicating that the two noises were mediated by different modes. **a**, In $\nu=2/3$, the C→N noise measured at $A_{NE}$ for -4nA≤$I_C$≤4nA with $t = 0.65$. **b**, In $\nu=2/3$, the C→N noise measured at $A_{NB}$ simultaneously with $A_{NE}$. **c**, In $\nu=1/3$, the C→N noise measured at $A_{NE}$ for -4nA≤$I_C$≤4nA with transmissions $t = 0.64$. **d**, In $\nu=1/3$, the C→N noise measured at $A_{NB}$ simultaneously with $A_{NE}$.

**Figure 6. | Additional test of neutral modes in a top-gated device.** We also investigated the presence of the neutral modes with a density-tunable device but without any QPC in the C→N configuration. **a**, The 2DEG was embedded within the light-blue line and the yellow pads are ohmic contacts (the source $S_C$ and the amplifier $A_{NE}$ to measure the C→N noise and ground $G$). Note that this configuration does not measure the edge contribution and the bulk contribution separately. The blue (red) arrows depict the charge (neutral) current. Top gates, colored in light green, for the left (TGL) and right halves (TGR) of the mesa tuned the local filling factors $\nu_L$ and $\nu_R$, respectively. We varied $\nu_R$ by the gate voltage $V_{TGR}$ on TGR and adjusted $\nu_L$ to be at 1 throughout the experiment. The defining gate (DG) in dark-green fully depleted the 2DEG underneath always. The density of 2DEG at $V_{TGR}=0$V was $1.3\times10^{11}$cm$^{-2}$. The interface of the left and right top gates was 3μm and the path length between the interface and the amplifier contact was 8 μm. **b**, At $B=7$T, driving $I_C=2.5$nA, the C→N noise as a function of $V_{TGR}$ was measured (blue dots) simultaneously with the bulk current $I_{bulk}$ (the Poissonian noise $2eI_{bulk}$ in red dots). The filing fractions ($\nu_R =1/3, 2/5, 3/5, 2/3$ and 1) were specified along the plot. The noise stayed finite at all the fillings except for $\nu_R=1$ and but gradually went down towards $\nu_R=1/3$, which is compatible with the observations shown above. Particularly, it is notable that, upon approaching $\nu_R=1/3$ after $\nu_R=2/5$, the bulk current vanished while the noise increased.

# Figure 1

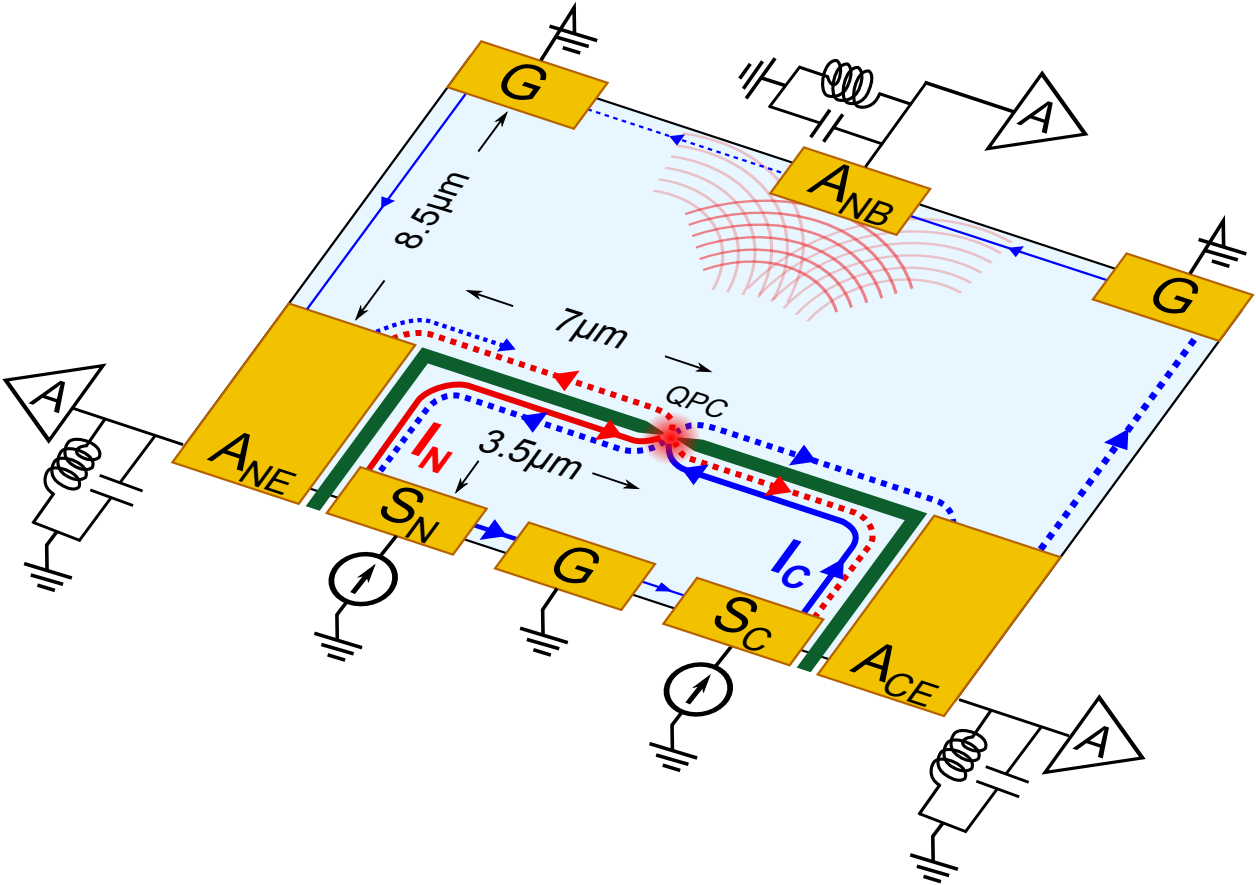

# Figure 2

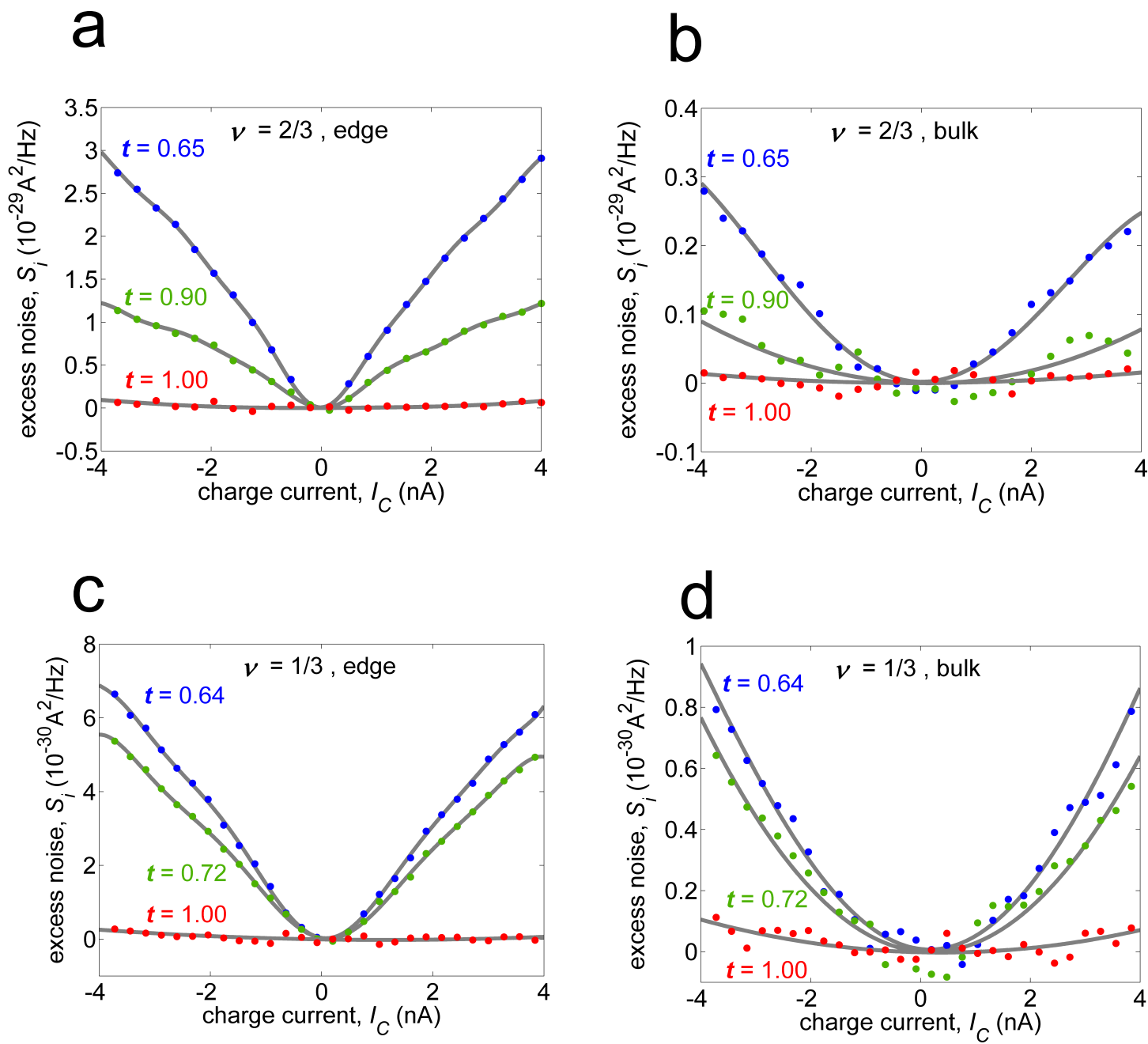

# Figure 3

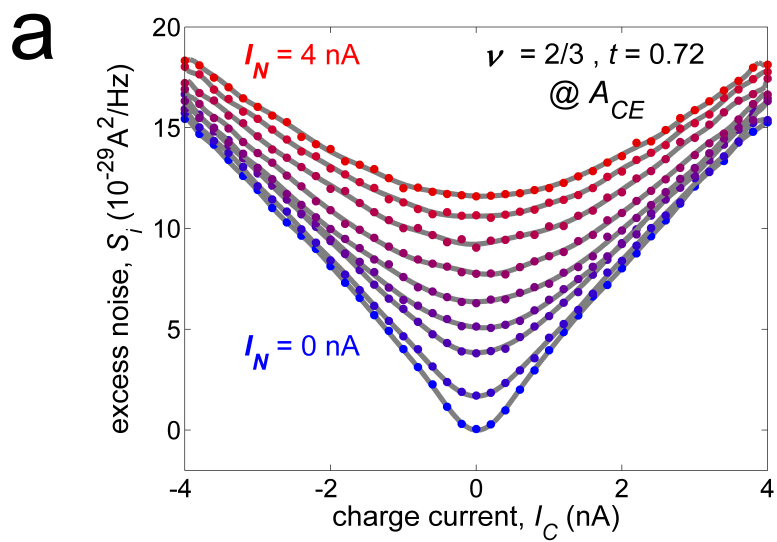

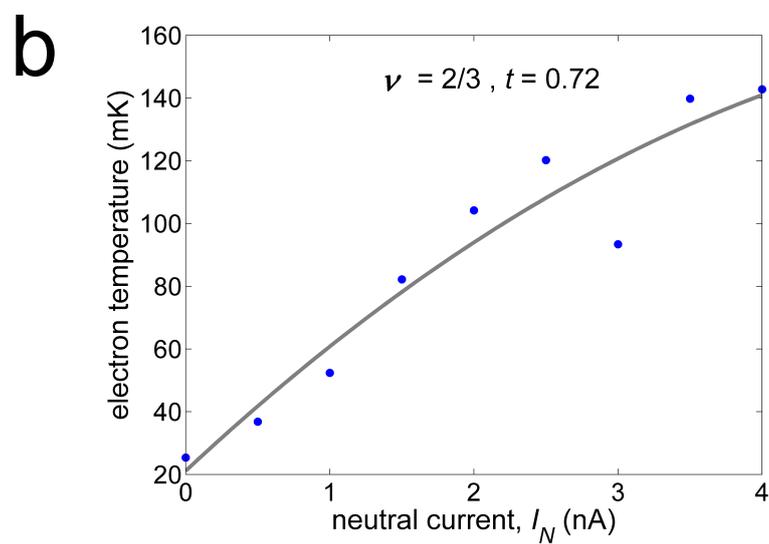

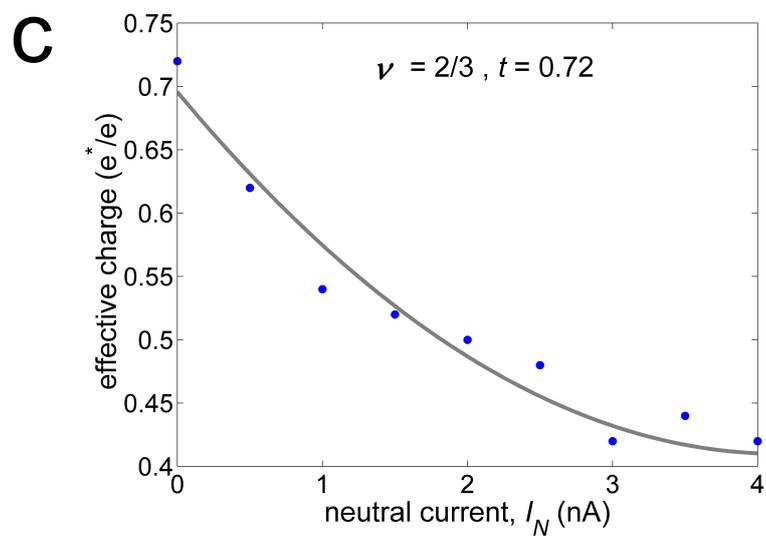

Figure 4

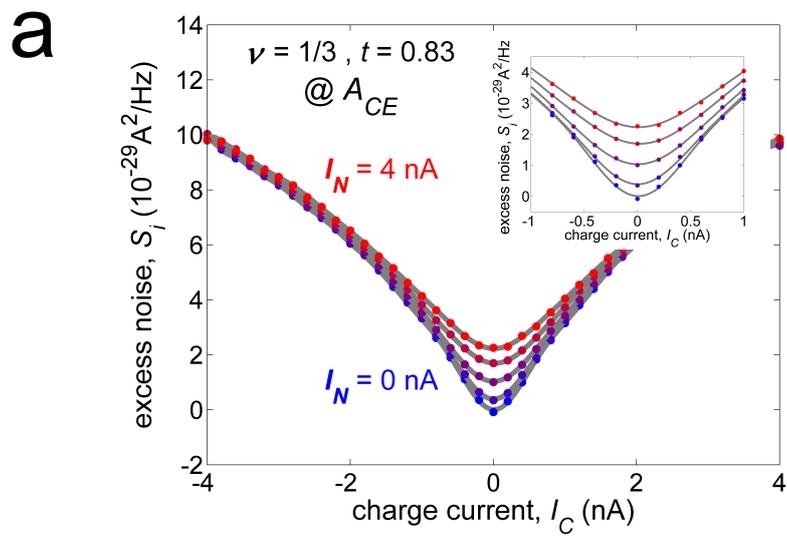

a

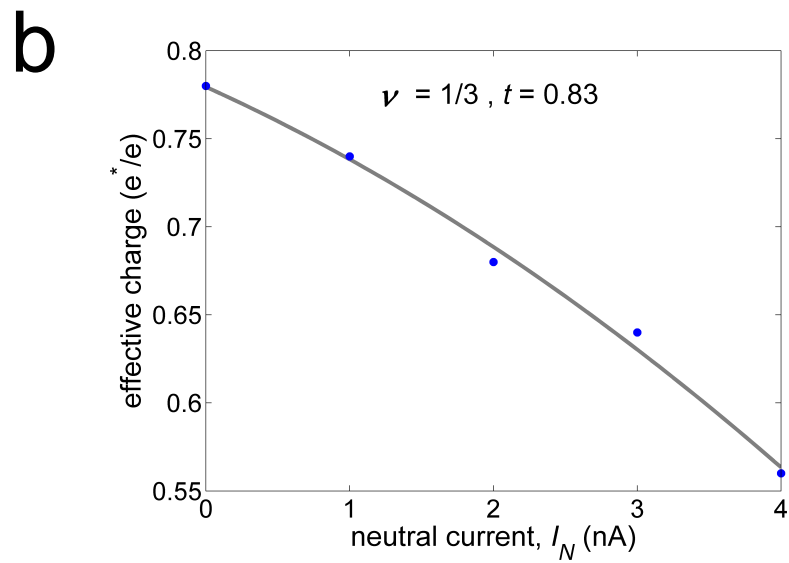

b

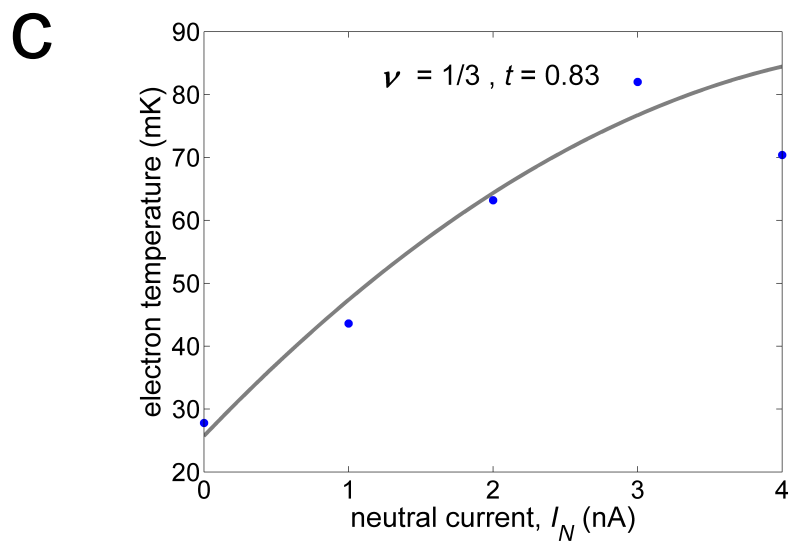

c

Figure 5

a
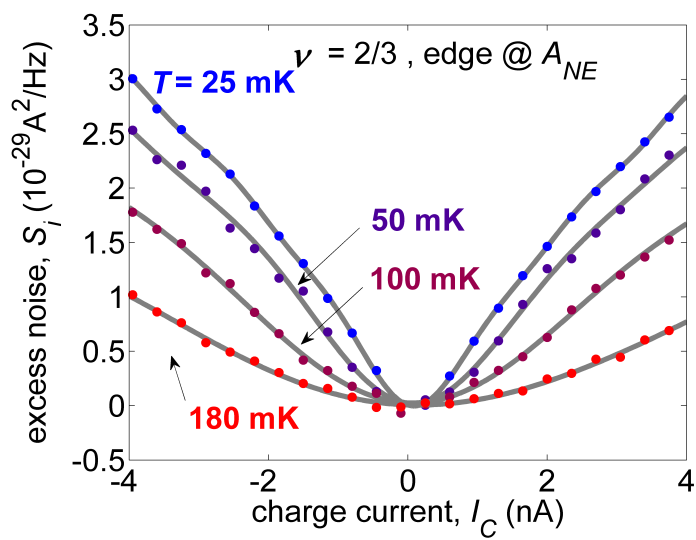

b
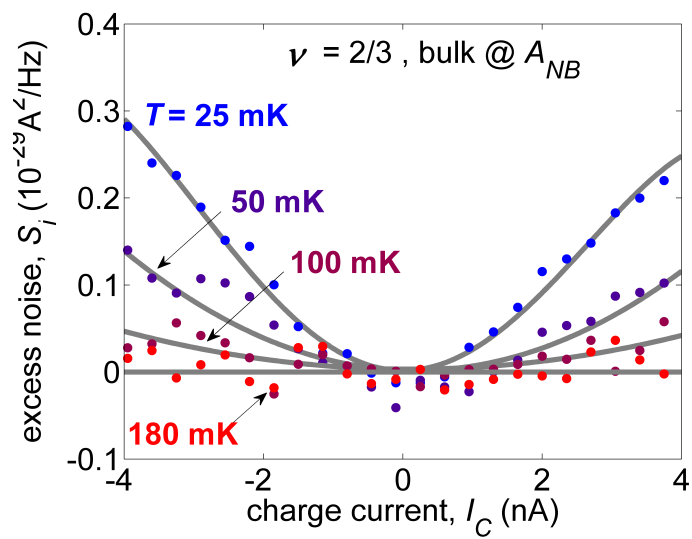

c
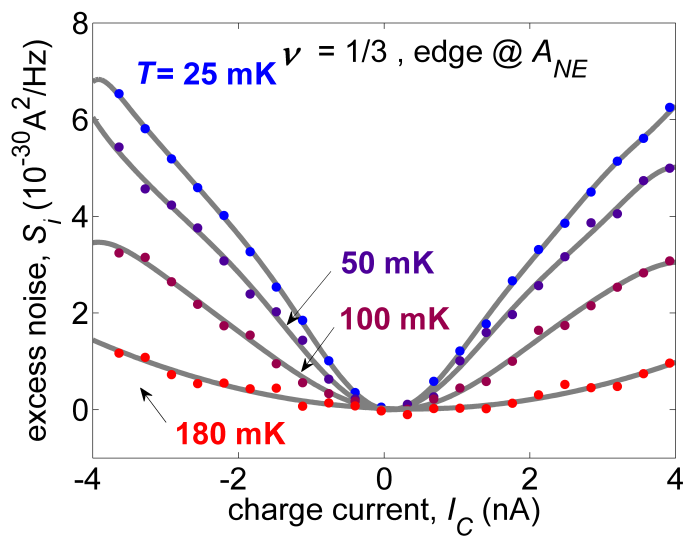

d
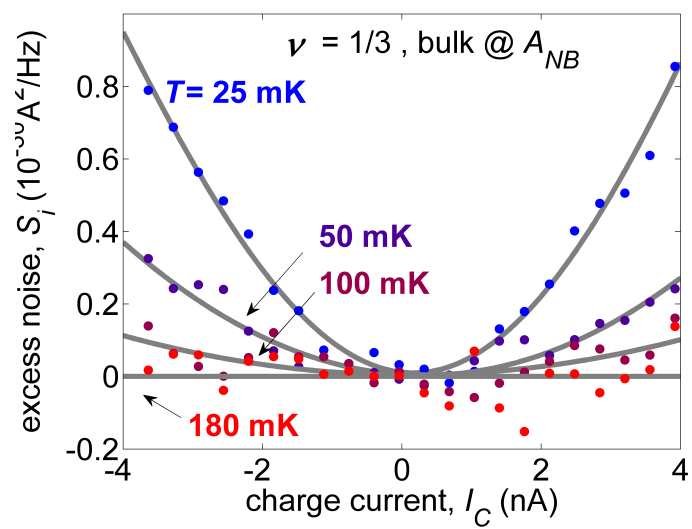

Figure 6

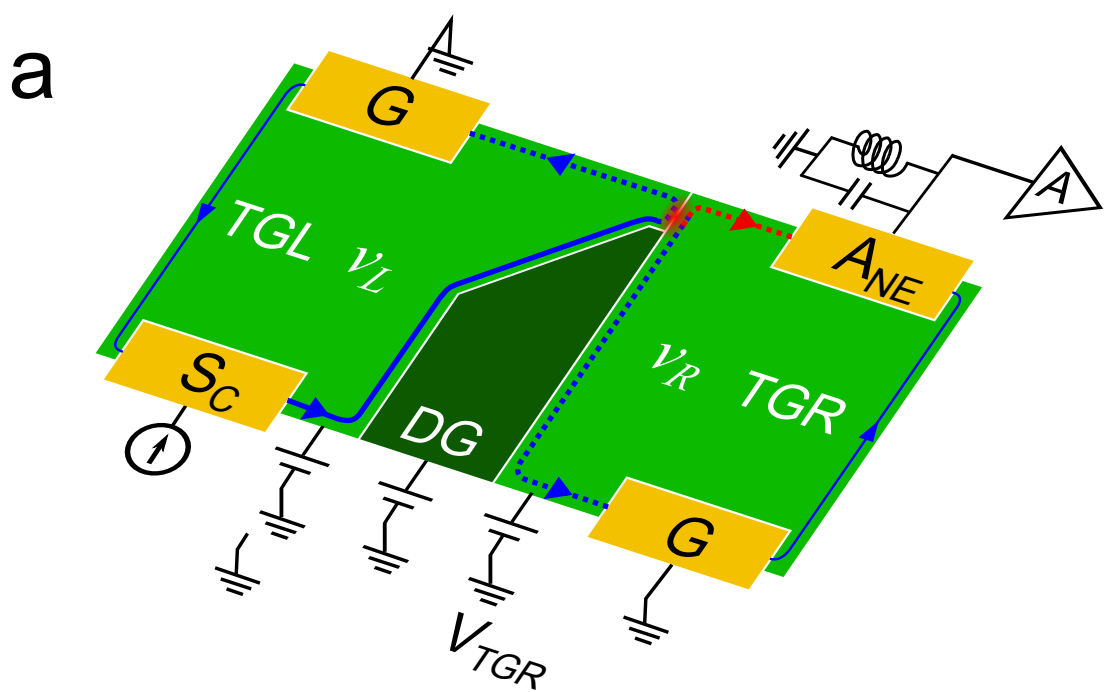

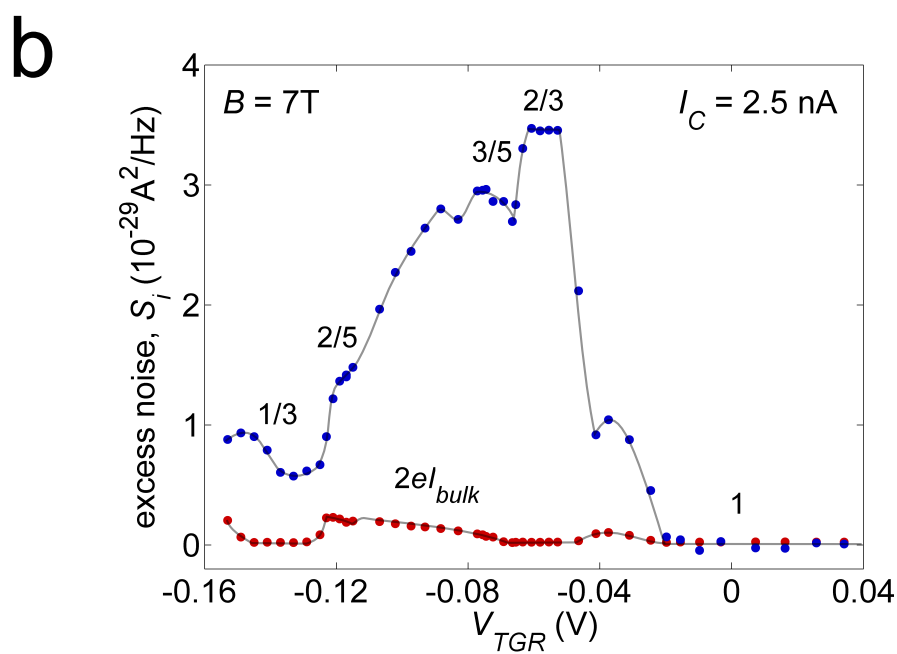

# Supplementary Materials

**Proliferation of neutral modes in fractional quantum Hall states**

Hiroyuki Inoue, Anna Grivnin, Yuval Ronen, Moty Heiblum, Vladimir Umansky and

Diana Mahalu

Braun Center for Submicron Research, Department of Condensed Matter Physics,

Weizmann Institute of Science, Rehovot 76100, Israel

**S1. Scanning electron micrograph images of the employed devices**

**S2. N→N noise at $\nu = 2/3$**

**S3. C→N noise at $\nu = 2/5$**

**S4. C→N noise at $\nu = 1$**

**S5. C→N noise with density-tunable device**

**S1. Scanning electron micrograph images of the employed devices**

In the present experiment, three kinds of configurations on different substrates were employed. The main samples were grown on a GaAs/AlGaAs heterostructure substrate with 2DEG embedded 116nm below the surface whose carrier density is $1.0 \times 10^{11} \mathrm{cm}^{-2}$ and the dark mobility is $5.1 \times 10^{6} \mathrm{cm}^{2}/\mathrm{Vs}$ at 4.2K (substrate 1). Two kinds of designs shown in Fig. S1a and S1b were made on this substrate. All the measurements described in the main manuscript were done on these samples. Let us denote the devices as A1 and B1 respectively (C→N noise measurement on A1 and C+N noise measurement on B1). Another kind of sample with top gates (shown in Fig. S1c) was fabricated on a substrate 2DEG with the carrier density of $1.3 \times 10^{11} \mathrm{cm}^{-2}$ (substrate 2). Let us denote it as device C2. The device C2 was utilized to take the data of Fig. 6. In this supplementary section (S6), another set of data was taken with a device of the design of Fig. S1b grown on a different substrate (substrate 3). Let us denote it as device B3. The substrate embedded 2DEG (113nm below the surface) with the carrier density of $1.2 \times 10^{11} \mathrm{cm}^{-2}$ and the dark mobility $4.2 \times 10^{6} \mathrm{cm}^{2}/\mathrm{Vs}$ at 4.2K.

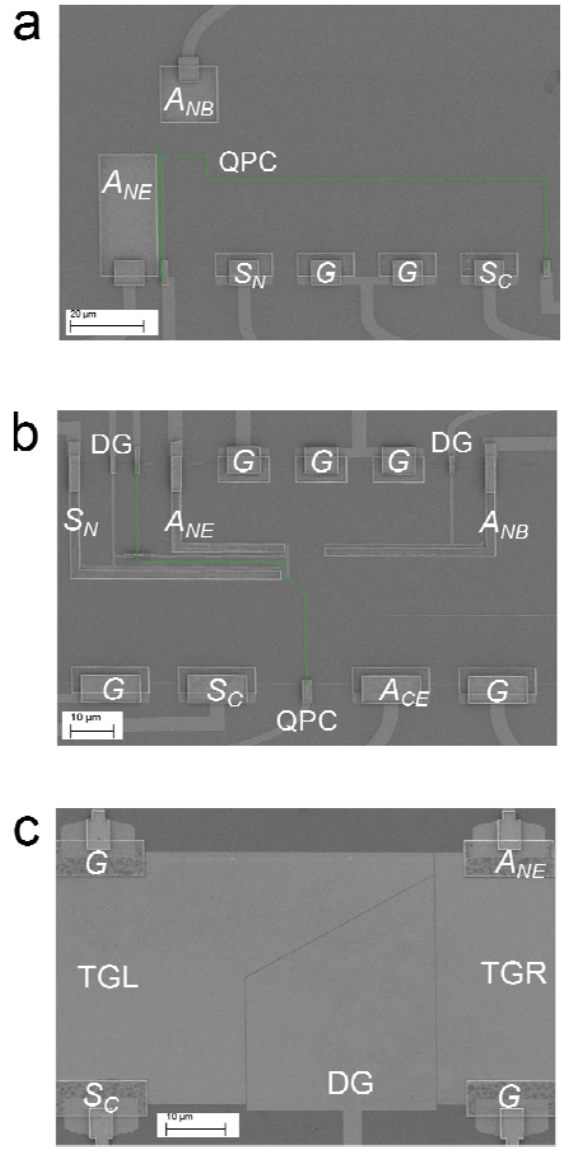

**Figure S1. | Scanning electron micrograph images of the employed devices.** We employed three kinds of designs from **a** to **c**. The configurations **a** and **b** are equivalent to the schematic illustration depicted in Fig. 1. They differ mainly in the distances between the contact $S_N$ and QPC, which were about 40μm in **a** and 3.5μm in **b**. For the detailed description of the samples, see the caption of Fig. 1 and Fig. 6a. The green high-lighted strips are the QPCs. The DGs always depleted the electron gas underneath.

## S2. N→N noise at ν = 2/3

In the device B3 the N→N noise at $\nu=2/3$ and $B=8.1$T was examined as a function of the QPC transmission probability $t$ (0.97, 0.78, 0.65, 0.49, 0) for $-4\text{nA} \leq I_N \leq 4\text{nA}$. Here, the N→N configuration means to source from $S_N$ and to measure at $A_{NE}$. The measured noise exhibited maximum when $t=0.97$ and monotonously decreased approximately as $t$ down to vanishing small noise.

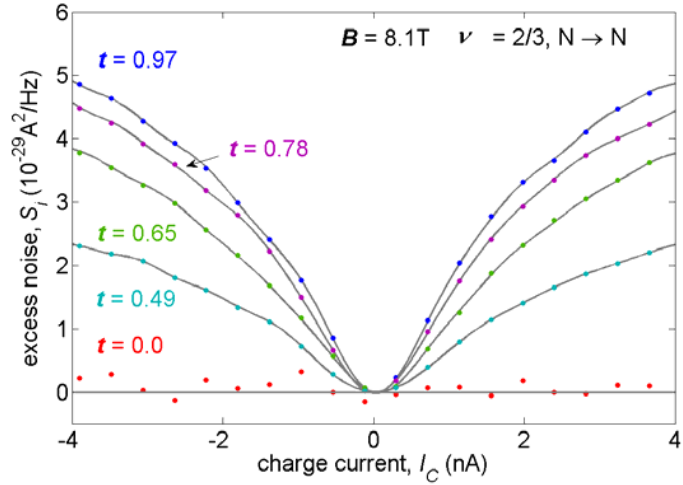

**Figure S2. | N→N noise at ν = 2/3.** Employing the sample with the density of $1.2\times10^{11}\text{cm}^{-2}$, the N→N configuration allows measuring noise of the neutral mode directly, without involving charge partitioning. We sourced current $I_C$ from $S_N$ and measure the noise reaching $A_{NE}$. Performing the measurements in $\nu=2/3$, for $-4\text{nA} \leq I_C \leq 4\text{nA}$ with QPC transmissions t=0.97, 0.78, 0.65, 0.49 and 0.0. As $|I_C|$ increases the noise also increases monotonically. The amplitude of the noise ascends with the transmission $t$.

**S3. C→N noise at ν = 2/5**

In the device B3 the C→N noise at ν=2/5 and $B$=11.2T was tested. Here, the edge of ν=2/5 possess two channels corresponding to the inner 2/5 channel and the outer 1/3 channel and each carries $I_C/6$ and $5I_C/6$ of the injected current $I_C$, which corresponds to the conductance of $G_0/15$ and $G_0/3$, respectively. Therefore, the transmission of the QPC as a function of the gate voltage exhibits a plateau corresponding to the full transmission of outer 1/3 channel. The noise measurement was tested for partitioning the inner 2/5 channel by 0.5, on the plateau of 1/3, and partitioning the outer 1/3 channel by 0.8. The excess noise vs effective current ($I_C/6$ and $5I_C/6$ for the inner and the outer channel respectively) is plotted. Finite noises were observed in all the cases including the case on the plateau which we do not expect to observe finite noise. This may be to do with a possible edge reconstruction therein leading to a partial mixing of two channels.

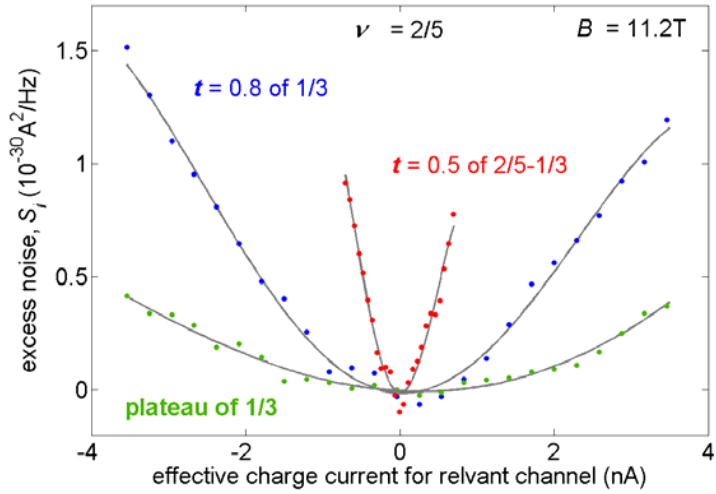

**Figure S3. | C→N noise at ν = 2/5.** C→N noise at ν=2/5 for several QPC transmissions. In this configuration there are two edge channels propagating one next to the other: An outer channel of ν=1/3 and an inner one with ν=2/5. The noise is measured when either one of the channels is partitioned and when the 1/3 channel is completely transmitted at the QPC and the 2/5 is reflected. The C→N noise is plotted as function of the effective current injected into the corresponding channel (The effective current injected into the 1/3 channel is $5/6×I_C$ and the one into the 2/5 channel is $1/6×I_C$).

## S4. C→N noise at ν = 1

In both devices A1 and B3, the C→N noise at ν=1 and B=4.4T and 5.5T were examined with at $A_{NE}$ and $A_{NB}$ for -4nA≤$I_C$≤4nA. In all cases, no appreciable noise were seen (all of them stayed below $0.5\times10^{-30}$ $A^2$/Hz). However, here we do not exclude the presence of neutral edge modes in integer fillings as reported in a previous study at ν=1 at a higher magnetic field. Since, at lower magnetic fields, it may not be favorable for the integer edges to reconstruct due to rather longer magnetic length being proportional to $B^{-0.5}$. Also, the structure of the edge can depend on details of the sample such as how the edge was defined (wet etch or gate defined), mobility of the 2DEG and etc.

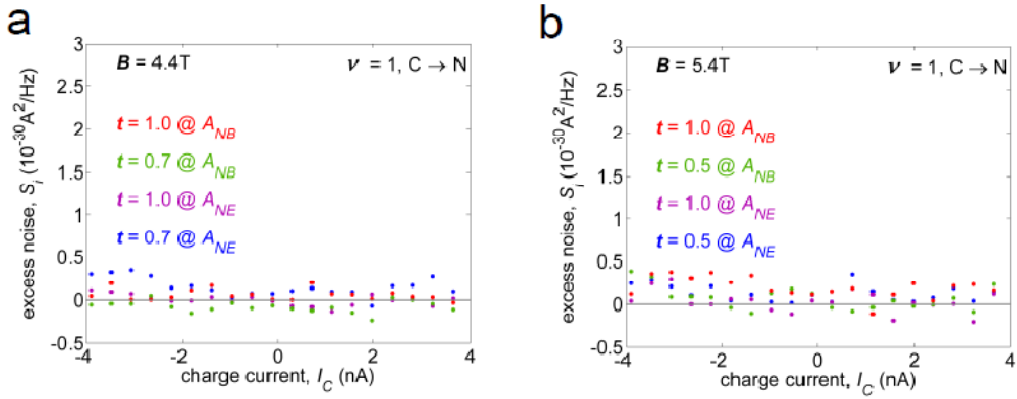

**Figure S4. | C→N noise at ν = 1.** C→N noise is measured in the upstream edge and over the bulk separately. **A,** In ν=1 (device A1) and B=4.4T the C→N noise at $A_{NB}$ and at $A_{NE}$ with transmissions $t$=1.0, 0.70 for -4nA≤$I_C$≤4nA. Unlike for the fractional fillings, as |$I_C$| increases, the noise remains below $0.5\times10^{-30}$ $A^2$/Hz. **b,** In ν=1 (device B3), B=5.4T the C→N noise at $A_{NB}$ and at $A_{NE}$ with transmissions $t$=1.0, 0.50 for -4nA≤$I_C$≤4nA. Similarly, the noise remains below $0.5\times10^{-30}$ $A^2$/Hz.

## S5. C→N noise with density-tunable device

With the device C2, the conductance from the left to the right half as a function of $V_{TGR}$ was measured with a standard lock-in technique (at 31Hz). Here, for example, being on the plateau of $v_r = 1/3$, a charge current of $I_C/3$ reaches the right half and builds the voltage of $I_C/3 \times 3G_0^{-1} = I_C G_0^{-1}$. Therefore, whenever the diagonal resistance $R_{xx}$ vanishes (forming a Hall plateau), it develops the same voltage of $I_C G_0^{-1}$, which is the unit of the color scale in the Fig. S5a. We repeated such conductance measurement at magnetic fields between 6T and 14T (Fig. S5a). The red strips correspond to the formation of the plateaus $v_R$=1/3, 2/5, 3/5, 2/3 and 1 from left to right.

Now, following the $v_R$=1/3, 2/3 and 1 strips, the C→N noise was then measured between $B$=7T and 14T (Fig. S5b). For each filling, the $V_{TGR}$ was adjusted to be on the middle of the plateau. While the noise at $v_R$ =1 stayed null, the noises in both fillings evolved in a similar fashion and the noise increased as the magnetic field was ramped down. The curves are merely guides for eyes. If we allow ourselves to attribute the observed noise solely to the upstream neutral edge modes because of their dominance over the bulk contribution, a lower density means a stronger disorder due to a weaker screening effect and hence may yield a bigger noise. Since the life time of upstream neutral mode at filling 2/3 was predicted to be proportional to disorder strength.

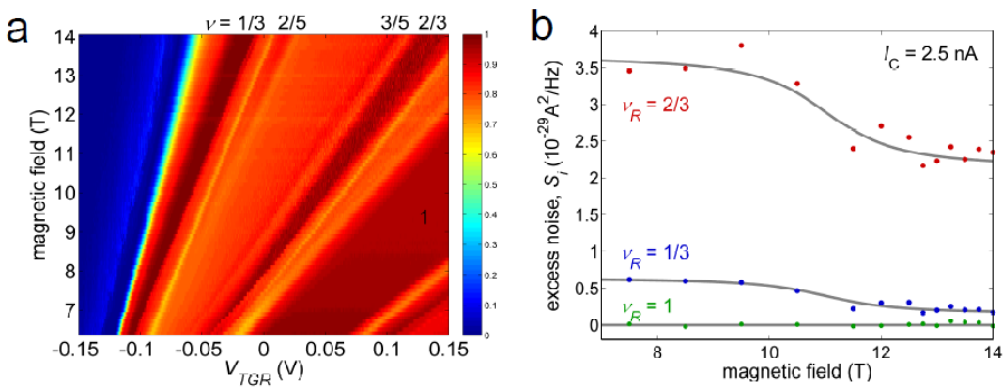

**Figure S5. | C→N noise with density-tunable device. a**, Fan diagram of the right half of the mesa is plotted $V_{TGR}$ vs $B$. The color scale is in the unit of $I_C G_0^{-1}$ and the red strips correspond to $v_R = 1/3, 2/5, 3/5, 2/3$ and 1 from left to right. **b**, The C→N noise for $v_R = 1/3, 2/3$ and 1 at various magnetic fields between 7T to 14T was measured. The noise in $v_R$ =1 stayed zero and the noise in the both fractions increased as the magnetic field was ramped down. The curves are just guide for eyes.